\documentstyle[12pt]{article}
\newcommand \kakko[1]{\left[ {#1} \right]}
\newcommand \ckakko[1]{\left\{ {#1} \right\}}
\newcommand \utilde[1]{\raisebox{-.8em}{$\tilde{~}$} \!\!\! {#1}}

\newcommand \ignore[1]{}

\newcommand \fra[2]{\displaystyle
{\frac{\textstyle {#1}}{\textstyle {#2}}}}

\makeatletter
\@addtoreset{equation}{section}
\makeatother

\begin{document}

\begin{flushright}
  UT-621 \\
  November 1992 \\
\end{flushright}
\vspace{24pt}
\begin{center}
\begin{large}
{\bf
The Ashtekar Formalism and WKB Wave Functions of N=1,2 Supergravities
}
\end{large}

\vspace{36pt}
        Takashi Sano\footnote{
        sano@tkyvax.phys.s.u-tokyo.ac.jp}

\vspace{6pt}

{\it Department of physics, University of Tokyo }\\
{\it Bunkyo-ku, Tokyo 113, Japan }

\vspace{48pt}

\underline{ABSTRACT}

\end{center}

\vspace{4cm}

The N=1,2 supergravities
with non-zero cosmological constants are investigated
in the Ashtekar formalism.
We solve the constraints of the N=1,2
supergravities semi-classically.
The resulting WKB wave functions
are expressed by exponentials
of supersymmetric-extended $SL(2,C)$ Chern-Simons
functional.

\vfill
\newpage

\ignore{}

\section{Introduction}

The quantization of the 4-dimensional gravity is the most interesting and
the most difficult problem in physics. Though there are
many attempts and efforts to solve this problem,
we have not yet obtained sufficient results.
It is due to the non-renormalizability
and the complexity of the Einstein-Hilbert action.
The path integral quantization, which is a powerful tool in the ordinary
quantum field theory, is not valid for the 4-dim quantum gravity, because
the Einstein-Hilbert action is bounded from neither above nor below owing to
the conformal factor of the metric  and it causes the divergence.
There are three typical attempts to overcome these difficulties:
the superstring theory, the lattice gravity, and the ADM canonical formalism.
Each approach has benefit and difficulty.

The superstring theory has been
the most promising one as the candidate of
the theory of everything.
The Einstein gravity can be derived as a low-energy
limit of the string theory.
Moreover since there is no ultraviolet divergence
in this theory,
it has more manageable perturbative behavior than the Einstein gravity.
In spite of much
success, the superstring theory has come to a deadlock because of the
mathematical difficulties. Even if we accept the string theory
 as a unified theory,
it seems difficult to use it for the quantization of the universe.
The conformal field theory,
which has its origin in the string theory, has developed rapidly and has been
applied to various situations. In particular the CFT has solved the
2-dimensional quantum gravity non-perturbatively.
But it seems difficult to apply similar method to the 4-dim gravity.

Recently the lattice gravity has progressed vigorously as a non-perturbative
approach to the 4-dim quantum gravity.
The examples are the Regge calculus and the Turaev-Viro theory, and so on.
We must ascertain whether all the matters which exist in our world can be
dealt with in lattice gravity.

The ADM  canonical formalism\cite{ADM},
in which we use the 3-dim metric as the
dynamical variables,
is used when we want to quantize
the gravity non-perturbatively. It is well known that the Hamiltonian
of the Einstein gravity reduces to a linear combination of the
 first class constraints. In the quantization by the ADM formalism,
 we consider these constraints
the quantum constraints and use them to select the physical states
from the space of functionals defined on the 3-dim metric.
 But unfortunately these constraints are non-polynomials of the
canonical variables and we can solve them
only when there are few degrees of freedom.
Moreover we have another difficult problem of operator ordering.
Since the Poisson bracket algebra of classical constraints represents
the diffeomorphism invariance of general relativity,
this and the commutator algebra of quantum constraints
must be isomorphic to each other. But unfortunately
there is no operator ordering in which required isomorphism is realized.

Recently a new formalism of the general relativity has been presented
by Ashtekar \cite{AA} \cite{AAA}. In this formalism,
we take the vierbein fields and the (anti-)self-dual part
of the Levi-Civita connection as fundamental variables.
It is shown that the constraints of the gravity
are simple polynomials of Ashtekar's canonical variables,
and we would expect that we can quantize the gravity non-perturbatively.
There have been some efforts already and a large class
of the physical states have been obtained\cite{JS}\cite{LS}\cite{HK}.

Some types of matters can be included in the Ashtekar formalism,
remaining the constraints to be polynomials of the
canonical variables. When we couple matters to gravity,
we must ascertain in each case whether the constraints are polynomials of the
canonical variables.
The typical and interesting examples which deal with matters
are the N=1,2 supergravities
\cite{CDJM}\cite{J}\cite{KS}. It is shown that the constraints are
polynomials of the canonical variables and the
Poisson brackets among the supersymmetry generators
are closed.

In this paper we discuss the Ashtekar formalism of the
N=1,2 supergravities and obtain the WKB wave functions.
In section 2 and section 3, the Ashtekar formalism of the
N=1,2 supergravities are given. These theories are
written by using the pure spin-connection formalism\cite{CDJM}.
The left- and the right-supersymmetry transformations are represented
in asymmetric ways.
The left-supersymmetry transformations
in both N=1 and 2 supergravity are parametrized
by usual left-handed spinor parameters. But the right-supersymmetry
transformation in N=1 case is realized by 1-form spinor parameter which
depends on field. In N=2 case, the right-supersymmetry transformation
is realized by 1-form spinor parameter and 1-form {\it bosonic}
parameter both of which depend on fields.
It is shown that the constraints
become polynomials of the canonical variables for N=1,2 supergravities,
with a slight modification for N=2 case.

In both N=1 and 2 case we calculate the commutators
of the supersymmetry transformations
on each field and find that they are not closed even on shell.
The cause of this feature is that the right-supersymmetry transformations
are realized by parameters which depend on fields.
We also calculate the Poisson algebras
of the supersymmetry generators and find that they are closed
among the generators of the local symmetries of the theories.
Therefore there is no inconsistency.

In section 4, we solve the constraints and obtain the WKB wave functions
of the supergravities with non-zero cosmological constants.
In \cite{HK} the WKB wave function of the pure gravity is obtained
when the cosmological term exists and
it has the form of exponential of the Chern-Simons functional.
The WKB wave functions of the N=1,2 supergravities have the forms
of exponentials of the N=1,2 supersymmetric-extended
Chern-Simons functional. Section 5 is devoted to the discussion.
The notations and the formulas used in this paper are given in the appendix.

Section 3 is based on the work with Dr. Kunitomo\cite{KS}.

\section{The Ashtekar Formalism of N=1 Supergravity}

In this section we consider the Ashtekar formalism of the N=1 supergravity.
The Ashtekar formalism of the N=1 supergravity
has been given first by Jacobson\cite{J} and reformulated in more
elegant form in ref\cite{CDJM}.

In this paper we use the method of the 2-form gravity\cite{CDJM}. We represent
the left- and the right-spinor indices as $A, B, C,\cdots$ and
$A', B', C',\cdots$, respectively. $e^{AA'}, \, \psi_{A},$ and
$ \psi_{A'}$ express the vierbein, the left- and the right-component
of gravitino. We define the 2-form fields
$\Sigma^{AB}$ and $ \chi^{A}$ as
\begin{eqnarray}
            \Sigma^{AB} &=& e^{A}_{A'}  \wedge e^{A'B}    \label{eq:Sigma}, \\
            \chi^{A}    &=& e^{A}_{A'}  \wedge \psi^{A'}. \label{eq:chi1}
\end{eqnarray}
The Lagrangian of N=1 supergravity with cosmological term
in the Ashtekar formalism
is given by \cite{KS}:
\begin{eqnarray}
       -i{\it{L}} &=& \Sigma_{AB}\wedge R^{AB}
                      +\chi^{A}\wedge D\psi_{A}
                      -\frac{1}{2}\Psi_{ABCD}\Sigma^{AB}\wedge\Sigma^{CD}
                      -\kappa_{ABC}\Sigma^{AB}\wedge\chi^{C} \nonumber \\
                  &&  - \frac{g^{2}}{6} \Sigma_{AB} \wedge \Sigma^{AB}
                      + \frac{1}{2} \lambda g \, \Sigma^{AB}
                      \wedge \psi_{A} \wedge \psi_{B}
                      -\frac{g}{6 \lambda} \chi_{A}
                      \wedge \chi^{A} \label{eq:LN1},
\end{eqnarray}
where $R_{AB}$ is the curvature of the anti-self-dual part of $ SO(3,1) $
connection $\omega_{AB}$, $ D $ is the covariant derivative
with respect to $\omega_{AB}$, and $ g $ and $\lambda $ are real constants.
$-\frac{g^{2}}{6}\Sigma_{AB}\wedge\Sigma^{AB}$ is cosmological term
with cosmological constant $\Lambda=g^{2}$.
$\Psi_{ABCD}$ and $\kappa_{ABC}$ are Lagrange multipliers which
require the following constraints:
\begin{eqnarray}
                \Sigma^{(AB} \wedge \Sigma^{CD)} &=& 0, \label{eq:cons11}\\
                \Sigma^{(AB} \wedge \chi^{C)} &=& 0, \label{eq:cons12}
\end{eqnarray}
where the indices between `(' and `)' are completely
symmetrized. These algebraic constraints
guarantee the decomposition (\ref{eq:Sigma}) and (\ref{eq:chi1}).

The Lagrangian (\ref{eq:LN1}) is invariant up to total derivatives under
the left- and the right local supersymmetries.
The left-supersymmetry transformation is given by:
\begin{eqnarray}
                \delta_{L}\Sigma^{AB} &=&
                -\chi^{(A} \,\epsilon^{B)}, \nonumber \\
                \delta_{L}\omega_{AB} &=&
                \lambda g \psi_{(A} \,\epsilon_{B)}, \nonumber \\
                \delta_{L}\psi_{A} &=&
                D \,\epsilon_{A}, \nonumber \\
                \delta_{L}\chi^{A} &=&
                \lambda g \Sigma^{AB}\,\epsilon_{B}, \nonumber \\
                \delta_{L}\kappa_{ABC} &=&
                -\Psi_{ABCD}\,\epsilon^{D}, \nonumber \\
                \delta_{L}\Psi_{ABCD} &=&
                -2\lambda g \,\kappa_{(ABC}\,\epsilon_{D)},
\end{eqnarray}
where $\epsilon_{A}$ is a left-handed spinor parameter.
The right-supersymmetry transformation is:
\begin{eqnarray}
                \delta_{R}\Sigma^{AB} &=&
                \psi^{(A}\wedge \eta^{B)}, \nonumber \\
                \delta_{R}\omega_{AB} &=&
                -\kappa_{ABC}\eta^{C}, \nonumber \\
                \delta_{R}\psi_{A} &=&
                \frac{g}{3\lambda}\eta_{A}, \nonumber \\
                \delta_{R}\chi_{A} &=&
                -D\eta_{A},\nonumber \\
                \delta_{R}\kappa_{ABC} &=& 0, \nonumber \\
                \delta_{R}\Psi_{ABCD} &=& 0,
\end{eqnarray}
where the parameter $\eta_{A}$ is
 1-form left-handed spinor parameter which
satisfies the following algebraic constraint:
\begin{equation}
                \Sigma^{(AB} \wedge \eta^{C)}=0. \label{eq:eta1}
\end{equation}
This condition can be solved on shell and $\eta^{A}$ is decomposed as
\begin{equation}
                \eta^{A}\sim e^{A}_{A'}\,\epsilon^{A'},
\end{equation}
where $\epsilon^{A'}$ is a right-handed spinor parameter.
Now we carry out the 3+1 decomposition of
 the Lagrangian (\ref{eq:LN1}).
We define variables $\tilde{\pi}^{iAB}$ and
$\tilde{\pi}^{iA}$ by
\begin{eqnarray}
                \tilde{\pi}^{iAB} &=&
                \frac{1}{2}\,\epsilon^{ijk}
                \Sigma^{AB}_{jk}, \label{eq:tpi}\\
                \tilde{\pi}^{iA} &=&
                \frac{1}{2}\,\epsilon^{ijk}\chi^{A}_{jk}.
\end{eqnarray}
Then the algebraic constraints (\ref{eq:cons11}), (\ref{eq:cons12})
are rewritten as
\begin{eqnarray}
              &&  \Sigma^{(AB}_{0i}\tilde{\pi}^{CD)i} = 0, \\
              &&  \Sigma^{(AB}_{0i}\tilde{\pi}^{C)i}
                +\tilde{\pi}^{i(AB}\chi^{C)}_{0i} = 0. \label{eq:App11}
\end{eqnarray}
As is shown in the appendix, we can solve these equations:
\begin{eqnarray}
                \Sigma^{AB}_{0i} &=& -\frac{1}{2}
                \,\epsilon_{ijk} \kakko{-i \;\utilde{N}
                \tilde{\pi}^{jA}_{C} \tilde{\pi}^{kCB}
                +2N^{j} \tilde{\pi}^{kAB} },  \label{eq:App12} \\
                \chi^{A}_{0i} &=& - \,\epsilon_{ijk} \kakko{
                -i \;\utilde{N} \tilde{\pi}^{jA}_{B} \tilde{\pi}^{kB}
                +N^{j}\tilde{\pi}^{kA}}
                +\,\epsilon_{ijk}\tilde{\pi}^{jA}_{B}
                \tilde{\pi}^{kBC}\utilde{M}_{C}, \label{eq:App13}
\end{eqnarray}
where $\utilde{M}_{A}$ is the spinor field of the weight -1, and
$\utilde{N}$ and $N^{i}$ correspond to the lapse function
and the shift vector in the ADM formalism, respectively.

The Lagrangian in the canonical form becomes
\begin{eqnarray}
                -iL &=& \tilde{\pi}^{iAB}\dot{\omega}_{iAB}
                +\tilde{\pi}^{iA}\dot{\psi}_{iA} \nonumber \\
                &&+\omega_{0AB}{\bf G}^{AB}-\psi_{0A}{\bf L}^{A}
                +\utilde{M}_{A}{\bf R}^{A}+\frac{1}{2}\,i\;\utilde{N}{\bf H}
                -N^{i}{\bf H}_{i}, \label{eq:gl1}
\end{eqnarray}
where $\omega_{0AB}$, $\psi_{0A}$, $\utilde{M}_{A}$,
$\utilde{N}$, and $N^{i}$ are the Lagrange multipliers.
The constraints are given by
\begin{eqnarray}
                {\bf G}^{AB} &=& D_{i}\tilde{\pi}^{iAB}
                -\psi^{(A}_{i}\tilde{\pi}^{B)i}, \label{eq:G1} \\
                {\bf L}^{A} &=& D_{i}\tilde{\pi}^{iA}
                -\lambda g \tilde{\pi}^{iAB}\psi_{Bi}, \label{eq:L1}\\
                {\bf R}^{A} &=& \tilde{\pi}^{iA}_{C}
                \tilde{\pi}^{jCB}
                \kakko{(D\psi_{B})_{ij}+\frac{g}{3\lambda}
                \,\epsilon_{ijk}\tilde{\pi}^{k}_{B}}, \label{eq:R1}\\
                {\bf H} &=& \tilde{\pi}^{iA}_{C}\tilde{\pi}^{jCB}
                \kakko{R_{ijAB}-\frac{g^{2}}{3}\,\epsilon_{ijk}
                \tilde{\pi}^{k}_{AB}+\lambda g
                \psi_{i(A}\psi_{B)j}} \nonumber \\
                &&+2\tilde{\pi}^{iA}_{B}\tilde{\pi}^{jB}
                \kakko{(D\psi_{A})_{ij}+\frac{g}{3\lambda}
                \,\epsilon_{ijk}\tilde{\pi}^{k}_{A}},\label{eq:H1} \\
                {\bf H}_{i} &=& \tilde{\pi}^{jAB}
                \kakko{R_{ijAB}-\frac{g^{2}}{3}\,\epsilon_{ijk}
                \tilde{\pi}^{k}_{AB}+\lambda g
                \psi_{i(A}\psi_{B)j}} \nonumber \\
                &&+\tilde{\pi}^{jA}\kakko{(D \psi_{A})_{ij}
                +\frac{g}{3\lambda}
                \,\epsilon_{ijk}\tilde{\pi}^{k}_{A}}. \label{eq:Hi1}
\end{eqnarray}
${\bf G}^{AB}$, ${\bf L}^{A}$, ${\bf R}^{A}$, ${\bf H}$, and ${\bf H}_{i}$
are the quantities with the weights +1, +1, +2, +2, and +1, respectively.

In this paper we always use the left derivatives for the fermionic fields.
The Poisson brackets between the canonical variables are
\begin{eqnarray}
                \ckakko{\omega_{iAB}(x,t), \,\tilde{\pi}^{jCD}(y,t)}
                &=& -i\delta^{j}_{i}\delta^{C}_{(A}\delta^{D}_{B)}
                \delta^{(3)}(x-y), \label{eq:PB11} \\
                \ckakko{\psi_{iA}(x,t), \,\tilde{\pi}^{jB}(y,t)}
                &=& -i\delta^{j}_{i}\delta^{B}_{A}
                \delta^{(3)}(x-y). \label{eq:PB12}
\end{eqnarray}
$-i$ in the r.h.s comes from the imaginary factor $-i$
in Lagrangian(\ref{eq:LN1}).
The constraints ${\bf G}^{AB}$, ${\bf L}^{A}$, ${\bf R}^{A}$, ${\bf H}$,
and ${\bf H}_{i}$ are the generators of
the local Lorentz transformation,
the left- and the right-supersymmetry transformations,
the time-reparametrization, and the 3-dim diffeomorphism.
These constraints are written in polynomials of the canonical variables.

Now let us calculate the commutators of the supersymmetry transformations
on each field and the Poisson brackets among the
 generators of the supersymmetries and compare the former and the latter.

First we calculate the commutators of the supersymmetry transformations
on each field.
These commutators must be closed among the local symmetries at least up to
the equations of motion. Hence these must have the forms:
\begin{eqnarray}
                &&\kakko{\delta_{1(L or R)},\,\delta_{2(L or R)}}\mbox{field}
                =(\sum_{\alpha \in S}
                \delta_{\alpha})\mbox{field}, \nonumber \\
                &&S :=\mbox{the set of the local symmetries
                of N=1 supergravity}.
\end{eqnarray}

 By using the explicit forms of the transformations, we can calculate the
commutators on each field. The left-left commutator is:
\begin{equation}
                \kakko{\delta_{L}(\,\epsilon_{1}),\,\delta_{L}(\,\epsilon_{2})}
                =\delta^{lL}(\Lambda)
\end{equation}
where $\delta^{lL}$ is the local Lorentz transformation and the parameter
${\Lambda^{A}}_{B}$ is given by
\begin{equation}
                {\Lambda^{A}}_{B}=\frac{1}{2}\lambda g (\,\epsilon^{A}_{1}
                \,\epsilon_{2B}+\,\epsilon_{1B}\,\epsilon^{A}_{2}).
\end{equation}
Here we use the equations of motion.
When we calculate the commutators which contain the right transformation,
we must note that the parameter
$\eta^{A}$ depends on $\Sigma_{AB}$ through (\ref{eq:eta1})
and $\eta^{A}$ must be transformed
such that the algebraic constraint (\ref{eq:eta1}) still holds after
the transformation:
\begin{equation}
                \delta_{L(R)}(\Sigma^{(AB}\wedge\eta^{C)})=0.\label{eq:etac1}
\end{equation}
The right-right commutator is then given by
\begin{equation}
                \kakko{\delta_{R}(\eta_{1}),\,\delta_{R}(\eta_{2})}
                =\delta_{R}(\eta_{3}),
\end{equation}
where $\eta_{3}$ is given by
\begin{equation}
                \eta^{A}_{3}=\delta_{R1}\eta^{A}_{2}
                -\delta_{R2}\eta^{A}_{1}.
\end{equation}
Here we used the equations of motion. The left-right commutator on the
fields except the auxiliary fields $\Psi_{ABCD}$ and $\kappa_{ABC}$ is
\begin{equation}
                \kakko{\delta_{L}(\,\epsilon),\,\delta_{R}(\eta)}
                =\delta^{Diff}(v)+\delta_{L}(\,\epsilon')+\delta_{R}(\eta')
                +\delta^{lL}(\Lambda), \label{eq:LR1}
\end{equation}
where $\delta^{Diff}$ represents the diffeomorphism. The parameters
appeared above are determined by:
\begin{eqnarray}
                i_{v}\Sigma^{AB} &=& \,\epsilon^{(A}\eta^{B)}, \nonumber \\
                \,\epsilon^{'A} &=& -i_{v}\psi^{A}, \nonumber \\
                \eta^{'A} &=& i_{v}\chi^{A}+\delta_{L}\eta^{A}, \nonumber \\
                \Lambda_{AB} &=& -i_{v}\omega_{AB}, \label{eq:par1}
\end{eqnarray}
where $i_{v}$ represents the interior product. By (\ref{eq:etac1})
and (\ref{eq:par1}) we obtain
\begin{equation}
                \Sigma^{(AB}\wedge\eta^{'C)}=i_{v}(\Sigma^{(AB}
                \wedge\chi^{C)}),
\end{equation}
and $\eta^{'A}$ satisfies (\ref{eq:eta1}) on shell.

The left-right commutators on the auxiliary fields are:
\begin{eqnarray}
                \kakko{\delta_{L}(\,\epsilon),\,\delta_{R}(\eta)}
                \Psi_{ABCD} &=& \cdots -i_{v}(D\Psi_{ABCD}
                +2\lambda g \kappa_{(ABC}\psi_{D)}), \nonumber \\
                \kakko{\delta_{L}(\,\epsilon),\,\delta_{R}(\eta)}
                \kappa_{ABC} &=& \cdots -i_{v}(D\kappa_{ABC}
                +\Psi_{ABCD}\psi^{D}).\label{eq:extra1}
\end{eqnarray}
where the parts $\cdots$ are the same as (\ref{eq:LR1}). The extra terms
in (\ref{eq:extra1}) don't disappear even on shell.
The cause
of this phenomenon seems to be that the parameter
$\eta^{A}$ depends on $\Sigma^{AB}$
through (\ref{eq:eta1}). To confirm this statement, next we calculate the
Poisson brackets of the constraints of the supersymmetry.
The reason why we calculate the Poisson brackets is as follows.
Since the constraints are
the generators of the local symmetries of the theory,
we can realize all the local symmetry transformations by constraints
and parameters which don't depend on fields.
If the Poisson bracket algebra of the constraints is closed,
we may consider that the algebra of the symmetries is closed.

The following smeared constraints are the generating functionals of the
symmetries:
\begin{equation}
                \begin{array}{lll}
                {\bf G}(\Lambda) = {\displaystyle \int}
                d^{3}x\Lambda_{AB}{\bf G}^{AB},&
                {\bf L}(\,\epsilon) = {\displaystyle \int}
                d^{3}x\,\epsilon_{A}{\bf L}^{A}, &
                {\bf R}(\eta) = {\displaystyle \int} d^{3}x\eta_{A}{\bf R}^{A},
                \\
                \\
                {\bf H}(N) = {\displaystyle \int} d^{3}xN{\bf H},&
                {\bf H}(\vec{N}) = {\displaystyle \int} d^{3}xN^{i}{\bf H}_{i},
                \end{array}
\end{equation}
where the parameters $\Lambda_{AB}$, $\epsilon_{A}$, $\eta_{A}$, $N$,
and $N^{i}$ are the quantities with the weights 0, 0, -1, -1, and 0,
respectively. Then the Poisson brackets among the supersymmetry
generators are:
\begin{eqnarray}
                \ckakko{{\bf L}(\,\epsilon_{1}), \,{\bf L}(\,\epsilon_{2})}
                &=& {\bf G}(\Lambda), \nonumber \\
                \Lambda_{AB} &=& -i\lambda g \,\epsilon_{1(A}\,\epsilon_{B)2},
\\
                \\
                \ckakko{{\bf L}(\,\epsilon), \, {\bf R}(\eta)}
                &=& {\bf H}(N)+{\bf H}(\vec{N}), \nonumber \\
                N &=& \frac{1}{2}i\,\epsilon_{A}\eta^{A}, \nonumber \\
                N^{i} &=& i\,\epsilon_{A}\eta_{B}\tilde{\pi}^{iAB}, \\
                \\
                \ckakko{{\bf R}(\eta_{1}), \, {\bf R}(\eta_{2})}
                &=& {\bf R}(\eta_{3}), \nonumber \\
                \eta_{3A} &=& -2i(\eta_{1(A}\eta_{B)2}\tilde{\pi}^{jB}_{C}
                \psi^{C}_{j}+\eta_{1(B}\eta_{C)2}\tilde{\pi}^{jB}_{A}
                \psi^{C}_{j}).
\end{eqnarray}
Consequently the Poisson brackets are closed among the local symmetries
in the theory. Therefore the algebra of the supersymmetry is closed.

\section{The Ashtekar Formalism of N=2 Supergravity}

In this section we discuss the Ashtekar formalism
of the N=2 supergravity\cite{KS}.
In the case of the N=2 supergravity the gravitinos form an $SU(2)$
doublet and $U(1)$ gauge field appears.

Let $\psi_{A}^{a}$ and
$ \psi_{A'}^{a}$ be the left- and the right-component
of gravitino, respectively.
The indices $a, b, c,\cdots$ are the internal $SU(2)$ indices.
We define 2-form field $ \chi^{A}_{a}$ as
\begin{equation}
            \chi^{A}_{a} = e^{A}_{A'}  \wedge \psi^{A'}_{a}. \label{eq:chi2}
\end{equation}
Now the chiral Lagrangian of the N=2 supergravity
is given by\cite{KS}:
\begin{eqnarray}
                -iL &=& \Sigma^{AB}\wedge R_{AB}
                +\chi^{A}_{a}\wedge D \psi^{a}_{A}
                -\frac{1}{2}\Psi_{ABCD}\Sigma^{AB}\wedge\Sigma^{CD}
                -\kappa^{a}_{ABC}\Sigma^{AB}\wedge\chi^{C}_{a} \nonumber \\
                &+& \phi_{AB}\hat{F}\wedge\Sigma^{AB}
                -\frac{1}{2}\phi_{AB}\phi_{CD}\Sigma^{AB}\wedge\Sigma^{CD}
                +\frac{1}{2}\phi_{AB}\chi^{A}_{a}\wedge\chi^{Ba}
                -\frac{1}{2}\hat{F}\wedge\hat{F},
\end{eqnarray}
and cosmological term and extra terms which are required when
we consider the theory with non-zero cosmological term are
\begin{eqnarray}
                -iL_{cosm}&=& g^{2}\Sigma_{AB}\wedge\Sigma^{AB}
                +g{(\tau^{3})_{a}}^{b}\Sigma^{AB}\wedge\psi^{a}_{A}
                \wedge\psi_{bB} \nonumber \\
                &-&\frac{1}{2}g{(\tau^{3})_{a}}^{b}\chi^{a}_{A}
                \wedge\chi^{A}_{b}
                -g{(\tau^{3})^{a}}_{b}\chi^{A}_{a}
                \wedge A \wedge\psi^{b}_{A},
\end{eqnarray}
where
$A$ is $U(1)$ gauge field,
$g$ is a real constant, and
$g^{2}\Sigma_{AB}\wedge\Sigma^{AB}$ is cosmological term with
cosmological constant $\Lambda=-6g^{2}$.

$\hat{F}$ is given by
\begin{equation}
                \hat{F}=dA-\frac{1}{2}\psi^{a}_{A}\wedge\psi^{A}_{a},
\end{equation}
and
\begin{equation}
                {(\tau^{3})}_{ab}=
                \left( \begin{array}{cc}
                1  &  0  \\
                0  &  -1
                \end{array}  \right).
\end{equation}
The Lagrange multipliers $\Psi_{ABCD}$ and $\kappa_{ABC}^{a}$
require the algebraic constraints
\begin{eqnarray}
                \Sigma^{(AB}\wedge\Sigma^{CD)} &=& 0, \label{eq:ac1}\\
                \Sigma^{(AB}\wedge\chi^{C)}_{a} &=& 0,\label{eq:ac2}
\end{eqnarray}
which guarantee the decomposition
(\ref{eq:Sigma}) and (\ref{eq:chi2}).

This action is invariant under the left- and right-supersymmetry
transformations up to total derivatives. The left-supersymmetry
transformation is:
\begin{eqnarray}
             \delta_{L}\Sigma^{AB}&=&-\chi^{(A}_{a}\,\epsilon^{B)a},
             \nonumber \\
             \delta_{L}\psi^{a}_{A}&=&D\,\epsilon^{a}_{A}
             -g{(\tau^{3})^{a}}_{b}A\,\epsilon^{b}_{A},
             \nonumber \\
             \delta_{L}\chi^{A}_{a}&=&(\widehat{F}
             -\phi_{CD}\Sigma^{CD})\,\epsilon^{A}_{a}+2g{(\tau^{3})^{b}}_{a}
             \Sigma^{AB}\,\epsilon_{bB},
             \nonumber \\
             \delta_{L}A&=&-\psi^{a}_{A}\,\epsilon^{A}_{a},
             \nonumber \\
             \delta_{L}\omega_{AB} &=& 2g{(\tau^{3})_{a}}^{b}
             \psi^{a}_{(A}\,\epsilon_{B)b},
             \nonumber \\
             \delta_{L}\Psi_{ABCD} &=& -4g{(\tau^{3})_{a}}^{b}
             \kappa^{a}_{(ABC}\,\epsilon_{D)b},
             \nonumber \\
             \delta_{L}\kappa^{a}_{ABC}&=&-\Psi_{ABCD}\,\epsilon^{aD}
             +3g{(\tau^{3})^{a}}_{b}\phi_{(AB}\,\epsilon^{b}_{C)},
             \nonumber \\
             \delta_{L}\phi_{AB}&=&\kappa^{a}_{ABC}\,\epsilon^{C}_{a},
\end{eqnarray}
where $\epsilon^{a}_{A}$ is a left-handed spinor parameter.

The right-supersymmetry transformation is:
\begin{eqnarray}
             \delta_{R}\Sigma^{AB}&=&\psi^{(A}_{a}\wedge\eta^{B)a},
             \nonumber \\
             \delta_{R}\psi^{a}_{A}&=&-\phi_{AB}\eta^{aB}
             -g{(\tau^{3})^{a}}_{b}\eta^{b}_{A},
             \nonumber \\
             \delta_{R}\chi^{A}_{a}&=&D\eta^{A}_{a}
             +g{(\tau^{3})_{a}}^{b}A\wedge\eta^{A}_{b}-\psi^{A}_{a}\wedge\xi,
             \nonumber \\
             \delta_{R}A&=&\xi,
             \nonumber \\
             \delta_{R}\omega_{AB}&=&\kappa^{a}_{ABC}\eta^{C}_{a}
             -\phi_{AB}\xi, \nonumber \\
             \delta_{R}\Psi_{ABCD} &=& 0, \nonumber \\
             \delta_{R}\kappa^{a}_{ABC} &=& 0, \nonumber \\
             \delta_{R}\phi_{AB} &=& 0,
\end{eqnarray}
where $\eta^{a}_{A}$ is a 1-form spinor
and $\xi$ is a {\it{bosonic}}
1-form which satisfy the constraints
\begin{eqnarray}
             \Sigma^{(AB}\wedge\eta^{C)}_{a}&=&0,\nonumber \\
             \Sigma^{AB}\wedge\xi&=&\chi^{a(A}\wedge\eta^{B)}_{a}.
             \label{eq:eta2}
\end{eqnarray}
We can solve these constraints on shell and obtain
\begin{eqnarray}
             \eta^{A}_{a}\sim e^{A}_{A'}\,\epsilon^{A'}_{a},
             \nonumber \\
             \xi\sim \psi^{A'}_{a}\,\epsilon^{a}_{A'},
\end {eqnarray}
where $\,\epsilon^{A'}_{a}$ is a right-handed spinor parameter.

Now let us consider the 3+1 decomposition. We define new variables:
\begin{eqnarray}
                \tilde{\pi}^{iAB} &=& \frac{1}{2}\,\epsilon^{ijk}
                \Sigma^{AB}_{jk}, \\
                \tilde{\pi}^{iA}_{a} &=& \frac{1}{2}
                \,\epsilon^{ijk}\chi^{A}_{ajk}, \\
                \tilde{\pi}^{i} &=& \phi_{AB}\tilde{\pi}^{iAB}
                -\frac{1}{2}\,\epsilon^{ijk}\hat{F}_{jk}.
\end{eqnarray}
Using these variables, (\ref{eq:ac1}) and (\ref{eq:ac2}) are rewritten as
\begin{eqnarray}
                \Sigma^{(AB}_{0i}\tilde{\pi}^{CD)i} &=& 0, \\
                \Sigma^{(AB}_{0i}\tilde{\pi}^{C)i}_{a}+
                \tilde{\pi}^{i(AB}\chi^{C)}_{a0i} &=& 0.\label{eq:D1}
\end{eqnarray}
These constraints can be solved as

\begin{eqnarray}
                \Sigma^{AB}_{0i} &=& -\frac{1}{2}
                \,\epsilon_{ijk} \kakko{-i \;\utilde{N}
                \tilde{\pi}^{jA}_{C} \tilde{\pi}^{kCB}
                +2N^{j} \tilde{\pi}^{kAB} },  \label{eq:App2} \\
                \chi^{A}_{a0i} &=& \,\epsilon_{ijk} \kakko{
                i \;\utilde{N} \tilde{\pi}^{jA}_{B} \tilde{\pi}^{kB}_{a}
                -N^{j}\tilde{\pi}^{kA}_{a}}
                +\,\epsilon_{ijk}\tilde{\pi}^{jA}_{B}
                \tilde{\pi}^{kBC}\utilde{M}_{Ca}, \label{eq:App3}
\end{eqnarray}
where $\utilde{M}_{Aa}$ is the fermionic field of the weight -1.

The Lagrangian is rewritten as
\begin{eqnarray}
                -iL&=&\tilde{\pi}^{iAB}\dot{\omega}_{iAB}
                +\tilde{\pi}^{iA}_{a}\dot{\psi}^{a}_{Ai}
                +\tilde{\pi}^{i}\dot{A}_{i} \nonumber \\
                &&+\omega_{0AB}{\bf G}^{AB}
                -\psi^{a}_{A0}{\bf L}^{A}_{a}
                +A_{0}{\bf G}+\frac{1}{2}i\utilde{N}{\bf H}-N^{i}{\bf H}_{i}
                -\utilde{M}^{a}_{A}{\bf R}^{A}_{a},
\end{eqnarray}
where $\omega_{0AB}$, $\psi^{a}_{0A}$, $A_{0}$, $\utilde{N}$, $N^{i}$
and $\utilde{M}^{a}_{A}$ are the Lagrange multipliers and the constraints are
\begin{eqnarray}
                {\bf G}^{AB} &=&
                D_{i}\tilde{\pi}^{iAB}+\tilde{\pi}^{i(A}_{a}\psi^{B)a}_{i},
                \label{eq:G2}\\
                {\bf L}^{A}_{a} &=& D_{i}\tilde{\pi}^{iA}_{a}+
                \tilde{\pi}^{i}\psi^{A}_{ia}
                +2g(\tau^{3})_{ab}\tilde{\pi}^{iAB}\psi^{b}_{iB}
                -g(\tau^{3})_{ab}\tilde{\pi}^{ibA}A_{i}, \\
                {\bf R}^{A}_{a} &=& \tilde{\pi}^{jA}_{B}\tilde{\pi}^{kCB}
                \kakko{ (D\psi_{aC})_{jk}+g(\tau^{3})_{ab}
                \,\epsilon_{ijk}\tilde{\pi}^{ib}_{C}
                -2g(\tau^{3})_{ab}A_{\left[j\right.}
                \psi^{b}_{\left.k\right]C}} \nonumber \\
                &&+\kakko{\,\epsilon_{ijk}\tilde{\pi}^{jAB}
                \tilde{\pi}^{k}_{aB}-h_{ij}\tilde{\pi}^{jA}_{a}}
                \tilde{\Pi}^{i}, \\
                {\bf G} &=& \partial_{i}\tilde{\pi}^{i}+g(\tau^{3})_{ab}
                \tilde{\pi}^{iaA}\psi^{b}_{iA},\\
                {\bf H} &=& \tilde{\pi}^{jA}_{C}\tilde{\pi}^{kCB}
                \kakko{R_{ABjk}+2g^{2}\,\epsilon_{ijk}\tilde{\pi}^{i}_{AB}
                -2g(\tau^{3})_{ab}\psi^{a}_{A\left[j\right.}
                \psi^{b}_{\left. k \right]B}} \nonumber \\
                && +2\tilde{\pi}^{jA}_{B}\tilde{\pi}^{kB}_{a}
                \kakko{ (D\psi^{a}_{A})_{jk}+g(\tau^{3})^{a}_{b}
                \,\epsilon_{ijk}\tilde{\pi}^{ib}_{A}
                -2g(\tau^{3})^{a}_{b}A_{\left[j\right.}
                \psi^{b}_{\left.k\right]A}} \nonumber \\
                && +\kakko{2h_{ij}\tilde{\pi}^{j}
                +\,\epsilon_{ijk}\tilde{\pi}^{ja}_{A}\tilde{\pi}^{kA}_{a}}
                \tilde{\Pi}^{i}, \\
                {\bf H}_{i} &=& \tilde{\pi}^{jAB}
                \kakko{R_{ABij}+2g^{2}\,\epsilon_{ijk}\tilde{\pi}^{k}_{AB}
                -2g(\tau^{3})_{ab}\psi^{a}_{A\left[i\right.}
                \psi^{b}_{\left. j \right]B}} \nonumber \\
                && +\tilde{\pi}^{jA}_{a}
                \kakko{ (D\psi^{a}_{A})_{ij}+g(\tau^{3})^{a}_{b}
                \,\epsilon_{ijk}\tilde{\pi}^{kb}_{A}
                -2g(\tau^{3})^{a}_{b}A_{\left[i\right.}
                \psi^{b}_{\left.j\right]A}} \nonumber \\
                && +\hat{F}_{ij}\tilde{\Pi}^{j},\label{eq:C2}
\end{eqnarray}
where $\tilde{\Pi}^{i}$ is defined by
\begin{equation}
                \tilde{\Pi}^{i}=\tilde{\pi}^{i}
                +\frac{1}{2}\,\epsilon^{ijk}\hat{F}_{jk},
\end{equation}
and $h_{ij}$ is the 3-dim space metric.
All of the above constraints are polynomials of the canonical variables
except ${\bf R}^{A}_{a}$ and $H$. The non-polynomial feature of these two
constraints result from that of $h_{ij}$. They become polynomials of the
canonical variables by multiplying them by $h:=\det h_{ij}$ because $h h_{ij}$
is a polynomial of the $\tilde{\pi}^{iAB}$.
${\bf G}^{AB}$, ${\bf L}^{A}_{a}$, ${\bf R}^{A}_{a}$, ${\bf G}$,
${\bf H}$, and ${\bf H}_{i}$ are the quantities with the weights
+1, +1, +2, +1, +2, and +1.

The Poisson brackets among the
canonical variables are
\begin{eqnarray}
                \ckakko{\omega_{iAB}(x,t),\,\tilde{\pi}^{jCD}(y,t)} &=&
                -i\delta_{i}^{j}\delta^{C}_{(A}\delta^{D}_{B)}
                \delta^{(3)}(x-y), \nonumber \\
                \ckakko{\psi^{a}_{iA}(x,t),\,\tilde{\pi}^{jB}_{b}(y,t)} &=&
                -i\delta_{i}^{j}\delta_{A}^{B}\delta_{b}^{a}
                \delta^{(3)}(x-y), \nonumber \\
                \ckakko{A_{i}(x,t),\,\tilde{\pi}^{j}(y,t)} &=&
                -i\delta_{i}^{j}\delta^{(3)}(x-y).\label{eq:PB2}
\end{eqnarray}

As in the case of the N=1 supergravity,
we must check whether the algebra of the supersymmetries is closed
among the local symmetries of the theory.
First we calculate the commutators of the supersymmetry transformations
on each field.
The left-left commutator is:
\begin{eqnarray}
                \ckakko{\delta_{L}(\,\epsilon_{1}),
\,\delta_{L}(\,\epsilon_{2})}
                &=&\delta^{lL}(\Lambda) +\delta^{U(1)}(\lambda), \nonumber \\
                \Lambda_{AB} &=& g(\tau^{3})^{a}_{b}(\,\epsilon_{1aA}
                \,\epsilon^{b}_{2B}+\,\epsilon^{b}_{1B}\,\epsilon_{2aA}),
\nonumber \\                \lambda &=&
-\,\epsilon^{a}_{1A}\,\epsilon^{A}_{2a},
\end{eqnarray}
where we use the equations of motion.

When we calculate the right-right and the left-right commutators
we must note that, since $\eta^{A}_{a}$ and $\xi$ depends on $\Sigma^{AB}$
and $\chi^{A}_{a}$ through the constraints
(\ref{eq:eta2}), $\eta^{A}_{a}$ and $\xi$ are also transformed by
supersymmetry transformations. These parameters must be transformed
as the following relations hold:
\begin{eqnarray}
             \delta_{L(R)}(\Sigma^{(AB}\wedge\eta^{C)}_{a})&=&0,\nonumber \\
             \delta_{L(R)}(\Sigma^{AB}\wedge\xi
             -\chi^{a(A}\wedge\eta^{B)}_{a})&=&0.
\end{eqnarray}
Using the above relations and the equations of motions,
 the right-right commutator for each field except $\omega_{AB}$, $\Psi_{ABCD}$,
and $\kappa^{a}_{ABC}$ is
\begin{equation}
                \kakko{\delta_{R}(\eta_{1},\,\xi_{1}),
                \,\delta_{R}(\eta_{2},\,\xi_{2})}=
                \delta_{R}(\eta_{3},\,\xi_{3})
                +\delta^{lL}(\Lambda)
                +\delta^{U(1)}(\lambda),
\end{equation}
where the parameters appearing above are determined by
\begin{eqnarray}
                \lambda\Sigma^{AB} &=& \eta^{a(A}_{1}\wedge\eta^{B)}_{2a},
                \nonumber \\
                \eta_{3a}^{A} &=& \delta_{R1}\eta^{A}_{2a}
                -\delta_{R2}\eta^{A}_{1a}+\lambda\psi^{A}_{a}, \nonumber \\
                \xi_{3} &=& \delta_{R1}\xi_{2}-\delta_{R2}\xi_{1}
                -d\lambda, \nonumber \\
                \Lambda_{AB} &=& -\lambda\phi_{AB}.
\end{eqnarray}
As can be seen easily, the following equations are obtained:
\begin{eqnarray}
                \Sigma^{(AB}\wedge\eta^{C)}_{3a} &=& 0,\nonumber \\
                \Sigma^{AB}\wedge\xi_{3} &=&
                \chi^{a(A}\wedge\eta^{B)}_{3a}
                +\lambda(D\Sigma^{AB}+\chi^{(A}_{a}\wedge\psi^{B)a}),
\end{eqnarray}
and hence $\eta^{A}_{3}$ and $\xi_{3}$ determine
a right-supersymmetry transformation
on shell.
The commutators for the remaining three fields are
\begin{eqnarray}
                \kakko{\delta_{R}(\eta_{1},\,\xi_{1}),
                \,\delta_{R}(\eta_{2},\,\xi_{2})}
                \omega_{AB} &=&
                \cdots +\lambda(D\phi_{AB}-\kappa^{a}_{ABC}\psi^{C}_{a}),
                \nonumber \\
                \kakko{\delta_{R}(\eta_{1},\,\xi_{1}),
                \,\delta_{R}(\eta_{2},\,\xi_{2})}
                \Psi_{ABCD} &=&
                \cdots -4\lambda\phi_{E(A}\Psi^{E}_{BCD)}, \nonumber \\
                \kakko{\delta_{R}(\eta_{1},\,\xi_{1}),
                \,\delta_{R}(\eta_{2},\,\xi_{2})}
                \kappa^{a}_{ABC} &=&
                \cdots -3\lambda\phi_{D(A}\kappa^{aD}_{BC)}
                -\lambda g (\tau^{3})^{a}_{b}\kappa^{b}_{ABC}.
\end{eqnarray}
The extra terms in these equations don't vanish even on shell. The same
situation happens in the left-right commutator.
The left-right commutator for each field except the auxiliary fields is
\begin{equation}
                \kakko{\delta_{L}(\,\epsilon),
                \,\delta_{R}(\eta,\,\xi)}=
                \delta^{Diff}(v)+\delta_{L}(\,\epsilon')
                +\delta_{R}(\eta',\,\xi')
                +\delta^{lL}(\Lambda)+\delta^{U(1)}(\lambda),
\end{equation}
and the parameters appearing above are determined by
\begin{eqnarray}
                i_{v}\Sigma^{AB} &=& \,\epsilon^{(A}_{a}\eta^{B)a},\nonumber \\
                \,\epsilon'^{a}_{A} &=& -i_{v}\psi^{a}_{A}, \nonumber \\
                \eta'^{A}_{a} &=& \delta_{L}\eta^{A}_{a}
                -\,\epsilon^{A}_{a}\xi-i_{v}\chi^{A}_{a}, \nonumber \\
                \xi' &=& \delta_{L}\xi-i_{v}(\hat{F}-\phi_{AB}\Sigma^{AB})
                -g(\tau^{3})_{a}^{b}\eta^{a}_{A}\,\epsilon^{A}_{b}, \nonumber
\\
                \Lambda_{AB} &=& -i_{v}\omega_{AB} \nonumber \\
                \lambda &=& -i_{v}A,
\end{eqnarray}
and as can be seen easily, one obtain the following equations:
\begin{eqnarray}
                \Sigma^{(AB}\wedge\eta'^{C)}_{a} &=& -i_{v}(
                \Sigma^{(AB}\wedge\chi^{C)}_{a}), \nonumber \\
                \Sigma^{AB}\wedge \xi' &=& \chi^{a(A}\wedge
                \eta'^{B)}_{a}-i_{v}\ckakko{(\hat{F}-\phi_{CD}\Sigma^{CD})
                \wedge\Sigma^{AB}+\frac{1}{2}\chi^{(A}_{a}\wedge\chi^{B)a}}.
\end{eqnarray}
Therefore $\eta'^{A}_{a}$ and $\xi'$ determine a right-supersymmetry
transformation on shell.
The commutators for the auxiliary fields are:
\begin{eqnarray}
                \kakko{\delta_{L}(\,\epsilon),
                \,\delta_{R}(\eta,\,\xi)}
                \Psi_{ABCD} &=&
                \cdots -i_{v}(D\Psi_{ABCD}+4g(\tau^{3})_{a}^{b}
                \kappa^{a}_{(ABC}\psi_{D)b}),\nonumber \\
                \kakko{\delta_{L}(\,\epsilon),
                \,\delta_{R}(\eta,\,\xi)}
                \kappa^{a}_{ABC} &=&
                \cdots -i_{v}(D\kappa^{a}_{ABC}
                -g(\tau^{3})^{a}_{b}A\kappa^{b}_{ABC} \nonumber \\
                &&\hspace{15mm}+\Psi_{ABCD}\psi^{aD}
                -3g(\tau^{3})^{a}_{b}\phi_{(AB}\psi^{b}_{C)}),\nonumber \\
                \kakko{\delta_{L}(\,\epsilon),
                \,\delta_{R}(\eta,\,\xi)}
                \phi_{AB} &=&
                \cdots -i_{v}(D\phi_{AB}-\kappa^{a}_{ABC}\psi^{C}_{a}).
\end{eqnarray}
The extra terms in these equations don't vanish even on shell.
As is in the case of the N=1 supergravity, the cause of this phenomenon
seems to be that the right-supersymmetry transformation is defined by
the parameters which depend on the fields.
 To ascertain directly that the algebra of the
 supersymmetry transformations is closed,
we must calculate the Poisson brackets
among the supersymmetry generators explicitly. We define the smeared
constraints as follows:
\begin{equation}
                \begin{array}{lll}
                {\bf G}(\Lambda) = {\displaystyle \int}
                d^{3} x \Lambda_{AB} {\bf G}^{AB}, &
                {\bf L}(\,\epsilon) = {\displaystyle \int}
                d^{3} x \,\epsilon^{a}_{A}
                {\bf L}^{A}_{a}, &
                {\bf R}(\eta) = {\displaystyle \int}
                d^{3} x \eta^{a}_{A} h {\bf R}^{A}_{a}, \\ \\
                G(\lambda) = {\displaystyle \int} d^{3} x \lambda {\bf G}, &
                {\bf H}(N) = {\displaystyle \int} d^{3} x N h^{2} {\bf H}, &
                {\bf H}(\vec{N}) = \frac{1}{2} {\displaystyle \int}
                d^{3} x N^{i} h^{2}
                {\bf H}_{i}, \\ \\
                \hat{\bf R}(\hat{\eta}) = {\displaystyle \int}
                d^{3} x {\hat{\eta}}^{a}_{A} h^{2}
                {\bf R}^{A}_{a},
                \end{array}
\end{equation}
where we define two smeared constraints for ${\bf R}^{A}_{a}$
for convenience of the calculation.
$\Lambda_{AB}$, $\epsilon^{a}_{A}$, $\eta^{a}_{A}$, $\lambda$,
$N$, $N^{i}$, and ${\hat{\eta}}^{a}_{A}$ are the quantities with
the weights 0, 0, -3, 0, -5, -4, and -5.

The Poisson brackets among the supersymmetry generators are:
\begin{eqnarray}
                \ckakko{{\bf L(\,\epsilon)},\,{\bf L(\,\epsilon')}} &=&
                G(\lambda)+{\bf G}(\Lambda), \nonumber \\
                \lambda &=& i\,\epsilon^{a}_{A}\,\epsilon'^{A}_{a},\nonumber \\
                \Lambda_{AB} &=& 2ig(\tau^{3})_{ab}\,\epsilon^{a}_{(A}
                \,\epsilon'^{b}_{B)}, \\
                \nonumber \\
                \ckakko{{\bf L}(\,\epsilon), \,\hat{{\bf R}}(\eta)} &=&
                {\bf H}(N)+\hat{{\bf R}}(\eta')+{\bf H}(\vec{N}),\nonumber \\
                N &=& -\frac{1}{2}i\,\epsilon^{A}_{a}\eta^{a}_{A},\nonumber \\
                {\eta'}^{a}_{A} &=& -\frac{1}{2h}i(2\,\epsilon^{bB}
                \eta^{a}_{A}\tilde{\pi}_{iBC}\tilde{\pi}^{iC}_{b}
                +\,\epsilon^{a}_{C}\eta_{Bb}\tilde{\pi}^{C}_{kA}
                \tilde{\pi}^{kbB}), \nonumber \\
                N^{i} &=& 2i\,\epsilon^{A}_{a}\eta^{aB}\tilde{\pi}^{i}_{AB},
                \\
                \nonumber \\
                \ckakko{{\bf R}(\eta_{1}),\,{\bf R}(\eta_{2})} &=&
                {\bf G}(\Lambda)+G(\lambda)+{\bf R}(\eta_{3}),\nonumber \\
                \Lambda_{AB} &=& 2i\eta_{1aC}\eta^{aC}_{2}h^{2}
                \tilde{\Pi}^{i}\tilde{\pi}_{iAB}, \nonumber \\
                \lambda &=& -4i\eta_{1aA}\eta^{aA}_{2}h^{3},\nonumber \\
                \eta^{A}_{3a} &=& 3i(\eta^{A}_{1a}\eta^{B}_{2b}
                +\eta^{B}_{1b}\eta^{A}_{2a})h\tilde{\pi}^{iC}_{B}\psi^{b}_{iC}
                \nonumber \\
                &&+i(\eta_{1aB}\eta_{2bC}+\eta_{1bC}\eta_{2aB})
                h\tilde{\pi}^{iAB}\psi^{bC}_{i},
\end{eqnarray}
and hence the Poisson bracket algebra of the supersymmetry generators
is closed among the generators of the local symmetries in the theory.



\section{The WKB Wave Functions}

In this section we consider the WKB wave functions of the N=1,2
supergravities with non-zero cosmological constants.
Similar WKB solution of the pure gravity is discussed in \cite{HK}.
Here we briefly review the case of the pure gravity.
The Ashtekar formalism of the pure gravity
with a non-zero cosmological constant is defined by the Lagrangian
\begin{equation}
                -iL=\Sigma_{AB} \wedge R^{AB}
                -\frac{\Lambda}{6}\Sigma_{AB}\wedge\Sigma^{AB}
                -\frac{1}{2}\Psi_{ABCD}\Sigma^{AB}\wedge\Sigma^{CD},
\end{equation}
and the constraints are
\begin{eqnarray}
                {\bf G}^{AB} &=& D_{i}\tilde{\pi}^{iAB},\nonumber \\
                {\bf H} &=& \tilde{\pi}^{iA}_{C}\tilde{\pi}^{jCB}
                (R_{ijAB}-\frac{\Lambda}{3}\,\epsilon_{ijk}
                \tilde{\pi}^{k}_{AB}), \nonumber \\
                {\bf H}_{i} &=& \tilde{\pi}^{jAB}(R_{ijAB}
                -\frac{\Lambda}{3}\,\epsilon_{ijk}\tilde{\pi}^{k}_{AB}),
                \label{eq:pgc}
\end{eqnarray}
where $\Lambda$ is cosmological constant and $\tilde{\pi}^{iAB}$
is defined by (\ref{eq:tpi}).
The Poisson brackets between the canonical variables are:
\begin{equation}
                \ckakko{\omega_{iAB}(x,t), \,\tilde{\pi}^{jCD}(y,t)}
                = -i \delta^{j}_{i} \delta^{C}_{(A} \delta^{D}_{B)}
                \delta^{(3)}(x-y).
\end{equation}
In the quantization, we replace the Poisson bracket by the canonical
commutation relation:
\begin{equation}
                \kakko{\omega_{iAB}(x,t), \,\tilde{\pi}^{jCD}(y,t)}
                = \delta^{j}_{i}\delta^{C}_{(A}\delta^{D}_{B)}
                \delta^{(3)}(x-y).
\end{equation}
We take the representation that $\omega_{iAB}$ is diagonalized:
\begin{equation}
                \tilde{\pi}^{iAB}=-\frac{\delta}{\delta\omega_{iAB}},
\end{equation}
and redefine the 1-form $\omega_{AB}$ as
\begin{equation}
                \omega_{AB}:=\omega_{iAB}dx^{i}, \label{eq:omega}
\end{equation}
where the index $i$ is the 3-dim space index; $i=1,2,3$.
Then we define the following functional:
\begin{eqnarray}
                \Phi\kakko{\omega_{iAB}}&=&
                \exp\kakko{-\frac{3}{2\Lambda}\int W}, \nonumber \\
                W &=& \omega_{AB} \wedge d \omega^{AB}
                +\frac{2}{3}\omega_{AC} \wedge \omega^{C}_{B}
                \wedge \omega^{AB}.\label{eq:WKB}
\end{eqnarray}
One can easily check that this functional vanishes by action of the constraints
(\ref{eq:pgc}) when the ordering of the operators is fixed
as in (\ref{eq:pgc}). This semi-classical
wave function has the form of exponential of the
Chern-Simons functional.

\vspace{5mm}
We can obtain similar semi-classical wave functions for the
N=1,2 supergravities. First we consider the case of the N=1 supergravity.
The canonical (anti-)commutation relations are:
\begin{eqnarray}
                \kakko{\omega_{iAB}(x,t), \,\tilde{\pi}^{jCD}(y,t)}
                &=& \delta^{j}_{i}\delta^{C}_{(A}\delta^{D}_{B)}
                \delta^{(3)}(x-y), \label{eq:CCR11} \\
                \ckakko{\psi_{iA}(x,t), \,\tilde{\pi}^{jB}(y,t)}
                &=& \delta^{j}_{i}\delta^{B}_{A}
                \delta^{(3)}(x-y). \label{eq:CCR12}
\end{eqnarray}
We take the representation in which $\omega_{iAB}$ and $\psi_{iA}$
are diagonalized:
\begin{eqnarray}
                \tilde{\pi}^{iAB} &=& -\frac{\delta}{\delta\omega_{iAB}},
                \nonumber \\
                \tilde{\pi}^{iA} &=& \frac{\delta}{\delta\psi_{iA}}.
\end{eqnarray}
We fix the ordering of the operators as in (\ref{eq:G1})-(\ref{eq:Hi1})
and redefine two 1-form fields $\omega_{AB}$ and $\psi_{A}$ as
\begin{eqnarray}
                \omega_{AB} &=& \omega_{iAB}dx^{i}, \nonumber \\
                \psi_{A} &=& \psi_{iA}dx^{i},
\end{eqnarray}
where $i=1,2,3$.
Then it can be easily seen that the following functional vanishes
by action of the constraints of the N=1 supergravity:
\begin{eqnarray}
                \Phi\kakko{\omega_{AB}, \psi_{A}} &=&
                \!\exp \kakko{-\frac{3}{2g^{2}}\int
                W_{N=1}  }, \label{eq:Phi}\\
                W_{N=1} &=& \omega_{AB}\wedge d \omega^{AB}
                \!+\frac{2}{3}\omega_{AC}\wedge\omega^{C}_{B}
                \wedge\omega^{AB}
                \!-\!\lambda g \psi^{A}\!\wedge
                \!D\psi_{A}.
\end{eqnarray}
As is expected, in the case of
the N=1 supergravity, the part of the Chern-Simons functional
in the case of the pure gravity is replaced by
supersymmetric-extended one; indeed, the functional $\int W_{N=1}$
is invariant under the local supersymmetry transformation
\begin{eqnarray}
                \delta\omega_{AB} &=& -\lambda g \,\,\epsilon_{(A}
                \psi_{B)}, \\
                \delta\psi_{A} &=& D\,\epsilon_{A},
\end{eqnarray}
where the covariant derivative $D$ is that corresponding
to the connection (\ref{eq:omega}).

Next we consider the case of the N=2 supergravity.
In quantization the Poisson brackets (\ref{eq:PB2})
are replaced by
\begin{eqnarray}
                \kakko{\omega_{iAB}(x,t),\,\tilde{\pi}^{jCD}(y,t)} &=&
                \delta_{i}^{j}\delta^{C}_{(A}\delta^{D}_{B)}
                \delta^{(3)}(x-y), \nonumber \\
                \ckakko{\psi^{a}_{iA}(x,t),\,\tilde{\pi}^{jB}_{b}(y,t)} &=&
                \delta_{i}^{j}\delta_{A}^{B}\delta_{b}^{a}
                \delta^{(3)}(x-y), \nonumber \\
                \kakko{A_{i}(x,t),\,\tilde{\pi}^{j}(y,t)} &=&
                \delta_{i}^{j}\delta^{(3)}(x-y).\label{eq:CCR2}
\end{eqnarray}
We take the representation in which $\omega_{iAB}$, $\psi^{a}_{iA}$,
and $A_{i}$ are diagonalized:
\begin{eqnarray}
                \tilde{\pi}^{iAB} &=& -\frac{\delta}{\delta\omega_{iAB}},
                \nonumber \\
                \tilde{\pi}^{iA}_{a} &=& \frac{\delta}{\delta\psi^{a}_{iA}},
                \nonumber \\
                \tilde{\pi}^{i} &=& -\frac{\delta}{\delta A_{i}}.
\end{eqnarray}
We fix the ordering of the operators as in (\ref{eq:G2})-(\ref{eq:C2}) and
redefine three 1-form fields $\omega_{AB}$, $\psi^{a}_{A}$,
and $A$ as follows:
\begin{eqnarray}
                \omega_{AB} &=& \omega_{iAB}dx^{i}, \nonumber \\
                \psi^{a}_{A} &=& \psi^{a}_{iA}dx^{i}, \nonumber \\
                A &=& A_{i}dx^{i}.
\end{eqnarray}
Then the WKB wave functional of the N=2 supergravity is given by
\begin{eqnarray}
                \Phi\kakko{\omega_{AB},\psi^{a}_{A},A} &=&
                \exp\kakko{\frac{1}{4g^{2}}\int W_{N=2}},\\
                W_{N=2} &=&
                 \omega_{AB} \wedge d \omega^{AB}
                +\frac{2}{3}\omega_{AC} \wedge \omega^{C}_{B}
                \wedge \omega^{AB}\nonumber \\
                &&-2g(\tau^{3})_{ab}\psi^{a}_{A}
                \wedge(D\psi^{bA}-g(\tau^{3})^{b}_{c}A\wedge\psi^{cA})
                +2g^{2}A\wedge dA.
\end{eqnarray}
As is in the case of the N=1 supergravity, the 3-form $W_{N=2}$
is the N=2 supersymmetric-extended Chern-Simons 3-form;
in fact, $\int W_{N=2}$ is invariant under the N=2
local supersymmetry transformation:
\begin{eqnarray}
                \delta\omega_{AB} &=& 2g(\tau^{3})_{ab}
                \,\epsilon^{a}_{(A}\psi^{b}_{B)}, \nonumber \\
                \delta\psi^{a}_{A} &=& D\,\epsilon^{a}_{A}
                -g(\tau^{3})^{a}_{b}A\,\epsilon^{b}_{A}, \nonumber \\
                \delta A &=& \,\epsilon^{a}_{A}\psi^{A}_{a}.
\end{eqnarray}

\section{Discussion}

In this paper we discussed the Ashtekar formalism of the N=1,2 supergravities
and their WKB wave functions. The constraints can be written in
polynomials of the canonical variables and the Poisson bracket algebras
of the supersymmetry generators are closed among the
generators of the local symmetries
of the theory. We solved the constraints semi-classically
and obtained the WKB wave functions of the N=1,2 supergravities. These
wave functions have the forms of exponentials of
the N=1,2 supersymmetric-extended Chern-Simons functional.

The Ashtekar formalism solve the difficulties which come from the
non-polynomial feature of the ADM formalism partly but it doesn't
solve them thoroughly in the following sense. Since the phase space of
the gravity is extended in the Ashtekar formalism,
we must set some ``reality conditions'' to
take out the net phase space from the extended one\cite{AA}.
These reality conditions are non-polynomials in general.
Furthermore the problem of the operator ordering is still left.

In this paper we use the representation in which the momentums
are replaced by the functional derivative operators.
This type of the quantization corresponds to the self-dual
representation\cite{JS}.
In ref\cite{LS}, the physical states are derived as the functionals defined
on the generalized knot space. This representation is called the
loop space representation. This and the self-dual representation
are dual each other.
It is an interesting problem
whether there exists a corresponding quantization program
for supergravities.
In the loop space representation of the pure gravity, the physical states
are related to the invariants of the knots. We may expect
that there are some invariants corresponding to the physical states of
supergravities. The WKB wave functions obtained in this paper
will be useful when we seek for the loop space representation of
supergravities.

\section*{Acknowledgement}

The author thanks H. Kunitomo, S. Iso, and J. Shiraishi for
helpful discussions and useful comments.


\section*{Appendix}

Here we sum up the notations and formulas used in this paper.
The space-time signature is $(-,+,+,+)$. We represent the 4-dim
space-time indices and the local Lorentz indices by $\mu, \nu, \rho,\cdots,$
and $a, b, c,\cdots,$ and the 3-dim space indices and the flat space indices by
$i, j, k,\cdots,$ and $I, J, K,\cdots,$ respectively.
The basis of
$SL(2,C)$ and $SU(2)$ spinors are
\begin{equation}
                \begin{array}{cc}
                \sigma^{0}_{AA'}=\fra{1}{\sqrt{2}}
                \left(  \begin{array}{cc}
                1 & 0 \\
                0 & 1 \end{array}
                \right),&
                \sigma^{1}_{AA'}=\fra{1}{\sqrt{2}}
                \left(  \begin{array}{cc}
                0 & 1 \\
                1 & 0 \end{array}
                \right),\\
                \sigma^{2}_{AA'}=\fra{1}{\sqrt{2}}
                \left(  \begin{array}{cc}
                0 & i \\
                -i & 0 \end{array}
                \right),&
                \sigma^{3}_{AA'}=\fra{1}{\sqrt{2}}
                \left(  \begin{array}{cc}
                1 & 0 \\
                0 & -1 \end{array}
                \right),
                \end{array}
\end{equation}
and
\begin{equation}
                \begin{array}{ccc}
                \tau^{1}_{AB}=
                2i\left(  \begin{array}{cc}
                1 & 0 \\
                0 & -1 \end{array}
                \right), &
                \tau^{2}_{AB}=
                2i\left(  \begin{array}{cc}
                i & 0 \\
                0 & i \end{array}
                \right), &
                \tau^{3}_{AB}=
                2i\left(  \begin{array}{cc}
                0 & -1 \\
                -1 & 0 \end{array}
                \right),
                \end{array}
\end{equation}
respectively. The $SO(3,1)$ vector $v_{a}$
is transformed into $SL(2,C)$ spinor as $v_{AA'}:=v_{a}\sigma^{a}_{AA'}$, and
the $SO(3)$ vector $u_{I}$ into $SU(2)$ spinor as $u_{AB}:=u_{I}\tau^{I}_{AB}$.
We define the anti-symmetric spinors by
\begin{equation}
                \,\epsilon_{AB}=\,\epsilon^{AB}
                =\,\epsilon_{A'B'}=\,\epsilon^{A'B'}
                =\left(  \begin{array}{cc}
                0 & 1 \\
                -1 & 0 \end{array} \right).
\end{equation}
The spinor indices can be raised and lowered according to the conventions
\begin{equation}
                \lambda^{A}=\,\epsilon^{AB}\lambda_{B},
                \,\,\,\lambda_{A}=\lambda^{B}\,\epsilon_{BA}.
\end{equation}

Taking an adequate gauge, we fix the form of the vierbein field as follows:
\begin{equation}
                {e_{\mu}}^{a}=
                \left(
                \begin{array}{cccc}
                N && N^{j}e_{j}^{I} & \\
                  &&& \\
                0 && e_{i}^{I} & \\
                  &&&
                \end{array}
                \right),
\end{equation}
where $N$ and $N^{i}$ are the lapse function and the shift vector,
respectively.
Defining the 3-dim space metric $h_{ij}$ by $h_{ij}=e^{I}_{i}e_{Ij}$,
the line-element of space-time is given by
\begin{equation}
                ds^{2}=-N^{2}dt^{2}+h_{ij}
                (N^{i}dt+dx^{i})(N^{j}dt+dx^{j}).
\end{equation}
Introducing the dual basis $e^{i}_{I}$ by $e^{i}_{I}e_{j}^{I}=\delta^{i}_{j}$,
we have
\begin{equation}
                \tilde{\pi}^{i}_{AB}=
                -\frac{1}{2}\sqrt{h}e^{i}_{I}\tau^{I}_{AB}
\end{equation}
by the straightforward calculation, where $h=\det h_{ij}$.
The following identities are obtained:
\begin{eqnarray}
                \tilde{\pi}^{i}_{AC}\tilde{\pi}^{jC}_{B} &=&
                hh^{ij}\,\epsilon_{AB}-\,\epsilon^{ijk}\tilde{\pi}_{kAB}, \\
                \tilde{\pi}_{iAC}\tilde{\pi}^{C}_{jB} &=&
                hh_{ij}\,\epsilon_{AB}-\,\epsilon_{ijk}h\tilde{\pi}^{k}_{AB},
\\
                hh_{ij} &=& \frac{1}{2}\tilde{\pi}^{AB}_{i}
                \tilde{\pi}_{jAB}, \\
                \tilde{\pi}_{iAB}\tilde{\pi}^{iCD} &=& 2h\delta^{C}_{(A}
                \delta^{D}_{B)}.
\end{eqnarray}
We also have
\begin{eqnarray}
                \Sigma^{AB}_{0i}&=&
                \frac{1}{2}Nie_{iI}\tau^{I AB}+\frac{1}{2}\,\epsilon_{ijk}
                N^{j}\sqrt{h}e^{k}_{I}\tau^{I AB} \\
                &=&-\utilde{N}i\tilde{\pi}^{AB}_{i}
                -\,\epsilon_{ijk}N^{j}\tilde{\pi}^{kAB},
\end{eqnarray}
where $\utilde{N}=N/\sqrt{h}$.
By the formula $\tilde{\pi}_{iAB}=-\frac{1}{2}\,\epsilon_{ijk}
\tilde{\pi}^{j}_{AC}\tilde{\pi}^{kC}_{B}$, finally we obtain
\begin{equation}
                \Sigma^{AB}_{0i}=
                -\frac{1}{2}\,\epsilon_{ijk}
                \kakko{-i\;\utilde{N}\tilde{\pi}^{jA}_{C}
                \tilde{\pi}^{kCB}+2N^{j}\tilde{\pi}^{kAB}}.
\end{equation}
We can ascertain that (\ref{eq:App13}) and (\ref{eq:App3}) are the solutions
 of (\ref{eq:App11}) and (\ref{eq:D1})
under (\ref{eq:App12}) and (\ref{eq:App2}), respectively. By counting of
the degrees of freedom, (\ref{eq:App13}) and (\ref{eq:App3}) are
just the general solutions.

\newpage


\begin{thebibliography}{99}
\bibitem{ADM} R. Arnowitt, S. Deser and C. W. Misner,
{\it in} Gravitation: an introduction to current research, ed.,
L. Witten (Wiley, New York,1962).
\bibitem{AA} A. Ashtekar, {\em Phys. Rev. lett.} {\bf 57} (1986) 2244;
{\em Phys. Rev.} {\bf D36} (1987) 1587.
\bibitem{AAA} A. Ashtekar, {\em "Lectures on Non-Perturbative Canonical
Gravity"}, World Scientific(1991).
\bibitem{JS} T. Jacobson and L. Smolin, {\em Nucl. Phys.} {\bf B299}
 (1988) 295.
\bibitem{LS} C. Rovelli and L. Smolin, {\em Nucl. Phys.} {\bf B331} (1990) 80.
\bibitem{CDJM} R. Capovilla, J. Dell, T. Jacobson and L. Mason,
{\em Class. Quantum Grav.} {\bf 8} (1991) 41.
\bibitem{J} T. Jacobson, {\em Class. Quantum Grav.} {\bf 5} (1988) 923.
\bibitem{KS} H. Kunitomo and T. Sano, {\it preprint}, OU-HET 167/UT-606,
{\em Int. J. Mod. Phys.}, to appear.
\bibitem{HK} H. Kodama, {\em Prog. Theor. Phys.} {\bf 80} (1988) 1024;
{\em Phys. Rev.} {\bf D42} (1990) 2548.
\end{thebibliography}
\end{document}